\documentclass[aps,nofootinbib,preprint,superscriptaddress,floatfix]{revtex4-1}
\usepackage{graphicx,color}
\usepackage{subfig}
\usepackage{amstext,amsmath,amssymb,amsfonts,amsthm}
\usepackage[format=plain,justification=centerlast]{caption}
\usepackage{comment}

\usepackage{mathrsfs}
\usepackage{hyperref}
\def\a{\alpha}
\def\b{\beta}
\def\g{\gamma}

\def\D{\Delta}

\def\no{\nonumber}
\def\be{\begin{equation}}
\def\ee{\end{equation}}
\def\ber{\begin{eqnarray}}
\def\eer{\end{eqnarray}}
\def\bwt{\begin{widetext}}
\def\ewt{\end{widetext}}

\begin{document}

\author{Carlos Villalpando}
\email{cvillalpando@ucol.mx}
\affiliation{Facultad de Ciencias - CUICBAS, Universidad de Colima, Colima, C.P. 28045, M\'exico}
\affiliation{Fermi National Accelerator Laboratory, Batavia, Illinois 60510, USA}

\author{Sujoy K. Modak}
\email{smodak@ucol.mx}
\affiliation{Facultad de Ciencias - CUICBAS, Universidad de Colima, Colima, C.P. 28045, M\'exico}


\title{Minimal length effect on the broadening of free wave-packets and its physical implications}

\begin{abstract}
We study the Generalized Uncertainty Principle (GUP) modified time evolution for the width of wave-packets for a scalar potential. Free particle case is solved exactly where the wave-packet broadening is modified by a coupling between the GUP parameter and higher order moments in the probability distribution in momentum space. We consider two popular forms of deformations widely used in the literature - one of which modifies the commutator with a quadratic term in momentum, while the other modifies it with terms both linear and quadratic in momentum. Unlike the standard case, satisfying Heisenberg uncertainty, here the GUP modified broadening rates, for both deformations, not only depend on the initial size (both in position and momentum space) of the wave-packet, but also on the initial probability distribution and momentum of the particle. The new rates of wave-packet broadening, for both situations, are modified by a handful of new terms - such as the \emph{skewness} and \emph{kurtosis} coefficients, as well as the (constant) momentum of the particle. Comparisons with the standard Heisenberg Uncertainty Principle (HUP)-based results show potentially measurable differences in the rates of free wave-packet broadening for physical systems such as the $C_{60}$ and $C_{176}$ molecules, and more so for large organic molecular wave-packets. In doing so, we open a path to scan the GUP parameter space by several orders of magnitude {\emph {inside}} the best existing upper bounds for both forms of GUP.
\end{abstract}

\maketitle
\tableofcontents

\section{Introduction}

One of the key features of Quantum Mechanics (QM) is the fact that it sets, by means of the Heisenberg Uncertainty Principle (HUP), a fundamental limit on the precise and simultaneous knowledge of two canonically conjugate dynamical variables for any quantum system. This, along with other fundamental principles, when put together, ensures the dispersion of free wave-packets through space in a manner that the width of the packet tends to always increase over time \cite{Messiah}. These insights are important to understand classical-quantum correspondence in general. For example, one can easily compute that the wave-packet corresponding to a free electron will disperse in space very rapidly and therefore the likelihood of pointing down a free electron to be present at  a specific point in space is negligible. Whereas, for a classical particle the wave-packet does not have a detectable dispersion in space over the age of the universe. 

The above features are of course very well in agreement  with our experiences which, nonetheless, are also verified in certain cases.  On the other hand there is a growing consensus that, inspired by certain quantum gravity theories, are advocating about the existence of a fundamental minimal length scale (at the Planck length). Among them, studies in string theory \cite{ST1, ST2, ST3, ST4, ST5, st-garay, st-castro}, Doubly Special Relativity \cite{DSR1, DSR2, DSR3, DSR4}, black hole physics \cite{BH1, BH2, BH3}, Loop Quantum Gravity (LQG) \cite{LQG1, LQG2}, non-commutative \textit{quantum} geometries \cite{NC1, NC2, NC3} and more general approaches concerning QM and General Relativity \cite{g1, g2, g3, Mangano, Nozari:2012gd, g4, g5, g6, Deriglazov:2016mhk, g7} manifest this existence of a minimal length by replacing the HUP by a Generalized Uncertainty Principle (GUP) whose exact form, however, often disagrees among various proposals (for a broad overview see \cite{BH1, sabine, tawfik, thesis, st-garay, gup-rev1, gup-rev2} and references therein). The GUP based approaches have a motivation to provide a short hand exercise in the search for quantum gravity effects, hypothesized to be realized in the form a minimal length, in low energy physics and, if it is indeed found then ask for an appropriate fundamental theory, from first principles, to explain this effective description of physical reality \cite{group-v2} (which may well be one of the existing theories - LQG or string theory or an entirely new theory).

One main focus of GUP based studies is to calculate the modified spectrum of different observables which can be useful to test the validity of the theory and in case no measurable differences are found it may still give bounds on the GUP parameters. Some of the studies in this line are reported in several works \cite{bounds}, and in fact a number of new experiments have been proposed \cite{exps} to measure these GUP contributions.

We, on the other hand, are opening a new avenue in this quest of understanding the fundamental insights that are brought in by the GUP modification (or the minimal length scale) on the {\em{wavefunction itself}} and thereby giving some new information on a distributional level. To do this we consider the wave-packet corresponding to a free particle which can give an account of the {\em{bare effect of minimal length scale}} on the otherwise very well understood situation. There exist preliminary works on the GUP effect on free particle wave-packets \cite{Nozari} but they were not developed enough to highlight the theoretical and experimental impact discussed here. Particularly, in this article, we present a detailed account  of the basic setting of wave-packet evolution both within the standard HUP setting (which is well-known) and within  the GUP framework (which is a new study). The GUP modification will be shown to imply a non-trivial distributional ramification on the rate and fundamental properties of broadening of the free wave-packets. We shall also compare two situations, i.e.,  GUP vs. HUP explicitly to clarify various outcomes, both mathematically and physically. There will be a considerable effort to estimate the time difference between the GUP and HUP broadening of the free wave-packets and the likelihood of experimentally detecting this departure. Interestingly, while doing so, we can also put some bounds on the GUP parameter and that will be an improvement of several orders of magnitude to the current best upper bounds that come from studying the spectrum of a number of observables. All these will be done using two most popular forms of the GUP given by the Ali-Das-Vagenas (ADV) form and the Kempf-Mann-Mangano (KMM) form. 

This paper is organized in the following manner: in the next section (\ref{hup}) we provide a review of the basic set up to derive the evolution law for the width of the free wave-packet. In section \ref{gee} we shall take first step to include the GUP effect by generalizing Ehrenfest's equations. Section \ref{gme} is used for the derivation of the governing equation for the spreading of free wave-packets in GUP scenario. Here we consider two popular forms of GUP, given by the ADV and KMM forms . The next section \ref{gup} is dedicated to solving these equations, exactly,  for the case of free particle case. Following section \ref{results} physically explain the new results. Moving on, in section \ref{tests} we shall elaborate on the possibility of testing our results within the present technology, for both forms of GUP. Finally, in section \ref{con} we conclude.

\section{The Motion and Spreading of Wave Packets}
\label{hup}

In this section we review the standard picture of wave packet broadening in quantum mechanics. This is a standard textbook exercise (see for example \cite{Messiah}), however, it is important to review it here for the sake of clarity and completeness of the paper.

\subsection{Ehrenfest's Theorem: a Classical Analogy}

In Quantum Mechanics, the fundamental principle that sets a limit in the precision to which one can simultaneously measure two given physical quantities, is the \textit{Heisenberg Uncertainty Principle} (HUP)  
\begin{equation}
\big[q_{i} , p_{j}\big] = i \hbar \delta_{ij} \hspace{0.5cm} , \hspace{0.5cm} i, j = 1,2, \ldots N,
\end{equation}
where $N$ is the number of spatial dimensions under consideration. This is equivalent to the uncertainty relationship between the position and momentum of a particle satisfying 
\begin{equation}
\Delta x \Delta p \geq \frac{\hbar}{2}. 
\end{equation}
We also have the identity applied to the time derivative of average (expectation) value of the observable
\begin{equation} \label{ident}
i\hbar \frac{d}{dt} \left< A \right> = \left< \big[ A, H \big] \right> + i\hbar \left< \frac{\partial A}{\partial t} \right>,
\end{equation}
where the observable  $A$ is understood as an self-adjoint operator and $H = H(q_{1}, \ldots , q_{N} ; p_{1}, \ldots, p_{N})$ is the system's Hamiltonian. Using this identity on coordinates of position and momentum, we obtain \emph{Ehrenfest's equations}
\begin{equation} \label{EE1}
\frac{d}{dt} \left< q_{i} \right> = \frac{1}{i \hbar} \left< \big[ q_{i}, H \big] \right> = \left< \frac{\partial H}{\partial p_{i}} \right>  \hspace{0.8cm}, \hspace{0.2cm} i = 1,2, \ldots, N
\end{equation}
and
\begin{equation} \label{EE2}
\frac{d}{dt} \left< p_{j} \right> = \frac{1}{i \hbar} \left< \big[ p_{j}, H \big] \right> = - \left< \frac{\partial H}{\partial q_{j}} \right>  \hspace{0.5cm}, \hspace{0.2cm} j = 1,2, \ldots, N.
\end{equation} 
which are deduced from the Ehrenfest's theorem. Notice that these equations are formally identical to Hamilton's equations in classical mechanics, although this formal analogy can only be rigorously made when the conditions
\bwt
\begin{equation*}
\left< \frac{\partial}{\partial p_{i}} H(q_{1}, \ldots, q_{N}; p_{1}, \ldots, p_{N}) \right> = \frac{\partial}{\partial p_{i}} H( \left< q_{1} \right>, \ldots, \left< q_{N} \right >; \left<p_{1}\right>, \ldots, \left<p_{N}\right> ) 
\end{equation*}
and
\begin{equation*}
\left< \frac{\partial}{\partial q_{j}} H(q_{1}, \ldots, q_{N}; p_{1}, \ldots, p_{N}) \right> = \frac{\partial}{\partial q_{j}} H( \left< q_{1} \right>, \ldots, \left< q_{N} \right >; \left<p_{1}\right>, \ldots, \left<p_{N}\right> ) 
\end{equation*} 
\ewt
are fulfilled. The above equations need not hold for an arbitrary potential, however, both of them hold perfectly up to the quadratic potential which then include the cases such as the free particle and the harmonic oscillator.

Now, let's consider a 1-dimensional wave packet $\Psi(q,t)$ with Hamiltonian
\begin{equation} \label{ham}
H = \frac{p^2}{2m} + V(q).
\end{equation}
In order to study the time evolution of the expectation values $ \left< q \right>$ and $\left< p \right>$ let's first define their mean-square deviations,
\begin{equation} \label{def}
\xi = (\Delta q)^2 = \left< q^2 \right> - \left< q \right>^2 \hspace{0.4cm} , \hspace{0.4cm} \eta = (\Delta p)^2 = \left< p^2 \right> - \left< p \right>^2 .
\end{equation}
Note that in the classical approximation $\Psi(q,t)$ represents a particle with position, momentum and energy given by
\begin{equation} \label{cl}
\begin{split}
q_{cl} = \left< q \right> \hspace{0.5cm} , \hspace{0.5cm} p_{cl} = \left< p \right>~~ \text{and}
~E_{cl} = \frac{\left< p \right>^2}{2m} + V \big(\left< q \right> \big).
\end{split}
\end{equation}
Now, let us define the quantity which tracks the difference
\begin{equation} \label{diff}
\varepsilon = \left< H \right> - E_{cl} = \frac{1}{2m} \eta + \left< V \right> - V_{cl}
\end{equation}
where $ V_{cl} = V \big(\left< q \right>\big)$. 

For the classical approximation to hold, we require the extension $\Delta q$ of the wave packet to remain small as compared to the characteristic distances of the problem under consideration, so that we can make the following Taylor expansions around $\left< q \right>$:
\begin{equation} \label{taylor}
\begin{split}
V(q) = V_{cl} + (q - \left< q \right>)V'_{cl} + \frac{1}{2}(q - \left< q \right>)^2 V_{cl}'' + \ldots \\
V'(q) = V_{cl}' + (q - \left< q \right>)V_{cl}'' + \frac{1}{2} (q - \left< q \right>)^2 V_{cl}''' + \ldots
\end{split}
\end{equation}
where $V_{cl}' = \frac{dV}{dq} \bigr\vert_{q = \left< q \right>}$. Using this expansion will guarantee the results are entirely general, i.e. valid for any $V$. Taking the expectation values of \eqref{taylor}, we obtain
\begin{equation} \label{taylorave}
\begin{split}
\left< V \right> = V_{cl} + \frac{1}{2} \xi V_{cl}'' + \ldots \\
\left< V' \right> = V'_{cl} + \frac{1}{2} \xi V_{cl}''' + \ldots
\end{split}
\end{equation}
By \eqref{EE1}, \eqref{EE2} and \eqref{ham}, we have
\begin{equation} \label{class}
\frac{d}{dt} \left< q \right> = \frac{\left< p \right>}{m} \hspace{0.5cm} , \hspace{0.5cm} \frac{d}{dt}\left< p \right> = - \left< V' \right>.
\end{equation}
Notice that, if we use $\left< V' \right> = V_{cl}'$ (\eqref{taylorave} up to first order), then equations \eqref{class} reduce to ``classical" equations of motion for the mean values $\left< q \right>$ and $\left< p \right>$. This result holds if $V(q)$ varies slowly over a distance $\sim \sqrt[]{\xi}$, so that the effect of $V'''$ and higher derivatives in \eqref{taylorave} is negligible. This condition holds trivially for the cases $V(q) = cq^2$ (harmonic oscillator) and $V(q) = 0$ (free particle), and for every $V(q)$ of at most order 2 in $q$. Assuming these conditions hold (i.e. series \eqref{taylorave} are rapidly converging), we have (see \eqref{diff})
\begin{equation} \label{eps}
\varepsilon \simeq \frac{1}{2m} (\eta + m V_{cl}''\xi) = constant
\end{equation}

\subsection{Deriving the Master Equation} \label{deriv}

We have described the motion of wave packets, by means of $\left< q \right>$ and $\left< p \right>$; now, in order to study the spreading of wave packets over time, we want to obtain functions $\xi (t)$, $\eta(t)$ (i.e. \textit{spread} in configuration and momentum space) explicitly. Notice that $\xi = \left< u \right>$, where $u = q^2 - \left< q \right>^2$, and $\left< q \right> = f(t)$, so applying identity \eqref{ident} to this operator yields
\begin{equation} \label{d}
\frac{d}{dt} \xi = \frac{1}{m} ( \left< pq + qp \right> - 2\left< p \right>\left< q \right> )
\end{equation}
Analogously, for the operator $d\xi / dt$, using again \eqref{ident} and \eqref{class} we obtain
\begin{equation} \label{d2}
\frac{d^2 \xi}{dt^2} = \frac{2\eta}{m^2}- \frac{1}{m}(\left< V'q + qV'  \right> -2 \left< q \right>\left< V' \right>).
\end{equation}
By using \eqref{taylor} in \eqref{d2}, we get the approximate equation
\begin{equation} \label{approx}
\frac{d^2 \xi}{dt^2} \simeq  \frac{2}{m^2} (\eta - m V_{cl}'' \xi ),
\end{equation}
and finally, taking \eqref{eps} into account, we can re-write it as
\begin{equation} \label{master}
\frac{d^2 \xi}{dt^2} \approx \frac{4}{m} (\varepsilon - V_{cl}'' \xi),
\end{equation}
which we refer here as the {\em{Master equation}}. Upon solving it, and knowing the deviations $\xi_{0}$, $\eta_{0}$, and $\dot{\xi_{0}} \equiv d\xi_{0} /dt$ at $t = t_{0}$, we obtain $\xi(t)$, the spread of the wave function over time in configuration space; $\eta (t)$ can then be found with \eqref{eps}, using the fact that $\varepsilon$ is constant.

Two interesting cases arise: the free particle and harmonic oscillator potential, in which the motion of the center of the packet is rigorously identical to that of a classical particle \cite{Messiah}. In the case of the free particle, $V = 0$, and thus from \eqref{eps} we have $\eta = 2m\varepsilon = \eta_{0}$, that is, $\eta = (\Delta p)^2$ remains constant. However, we have rigorously $d^2\xi/dt^2 = 2\eta_{0}/m^2$ and thus
\begin{equation} \label{free}
\xi (t) = \xi_{0} + \dot{\xi_{0}}t + \frac{\eta_{0}}{m^2}t^2.
\end{equation}
This result tells us that the free wave packet \textit{spreads} indefinitely, as is well known, {\em so this sets a limit for the time interval during which the classical-particle analogy holds}. If we have $\dot{\xi_{0}} = 0$ (e.g., the packet has the {\em{minimum}} width at $t_{0}$, so that, $\xi_{0}\eta_{0} = \frac{1}{2}\hbar^2$) then \eqref{free} is simplified to $\xi = \xi_{0} + \eta_{0}t^2/m^2$ or, equivalently,
\begin{eqnarray}
\Delta q (t) = \sqrt{\xi} (t) = \left[ (\Delta q_{0})^2 + \left( \frac{\Delta p_{0} t}{m} \right)^2 \right]^{1/2},
\label{freeex}
\end{eqnarray}
where $\Delta q_{0}$ and $\Delta p_{0}$ are the initial uncertainty in position and momentum space corresponding to the minimum wave-packet. This is a truly remarkable equation and fundamental to our physical understanding of quantum theory which explains why we cannot see an electron as a localized object and why classical objects are seem to be localized forever. Take for instance the case of free electron - the second term in \eqref{freeex} increases with time as $t^2$ and matches the initial width in time $t=\frac{2\pi (\Delta q_0)^2}{c \lambda_e}$ (by using the minimum wave-packet uncertainty relation $\Delta q_0 \Delta p_0 = \hbar/2$ and the definition for the Compton wavelength for electron). Using $\lambda_e = 2.4 \times 10^{-12}~m$ and initial width $\Delta q_0 \simeq 10^{-10}~m$ we get the time it takes for the second term in \eqref{freeex} to equate the first term is $t\sim 10^{-16}~s$. This is why it is hard to detect electron as a localized object confined to a small space - the wave-packet gets quickly delocalized. On the other hand for most of the classical objects this time is more than the age of the universe.


\section{The Generalized Ehrenfest Equations}
\label{gee}

The general form of the GUP commutator we are considering here is given by \cite{our-gup, our-gup2}
\begin{equation} \label{GUPgen}
\big[ q_{i}, p_{j} \big] = i\hbar \Big\{ \delta_{ij} - \alpha \big(p\delta_{ij} + \frac{p_{i} p_{j}}{p} \big) + \beta^2 (p^2 \delta_{ij} + 3p_{i}p_{j}) \Big\}.
\end{equation}
This relationship is valid for the non-relativistic quantum mechanical context where the coordinates are usually identified as the Cartesian coordinates (which is true even with standard commutator without GUP modifications \cite{Messiah}). This leads to the Generalized Ehrenfest Equations
\ber 
\label{GEEgen1}
\frac{d}{dt} \langle q_{i} \rangle &=& \sum_{j = 1}^{N} \left\{ \delta_{ij} \left< \frac{\partial H}{\partial p_{j}} \right> - \alpha \left( \delta_{ij} \left< p \hspace{0.08cm} \frac{\partial H}{\partial p_{j}} \right> + \left< \frac{p_{i} p_{j}}{p} \frac{\partial H}{\partial p_{j}} \right> \right) \right. \notag \\
&& \left.+ \beta^2 \left( \delta_{ij} \left< p^2 \frac{\partial H}{\partial p_{j}} \right> + 3\left< p_{i} p_{j} \frac{\partial H}{\partial p_{j}} \right> \right) \right\}
\eer

\ber
\label{GEEgen2}
\frac{d}{dt} \left< p_{i} \right> &=& \sum_{j = 1}^{N} \left\{ - \delta_{ij} \left< \frac{\partial H}{\partial q_{j}} \right> + \alpha \left( \delta_{ij} \left< p \hspace{0.08cm} \frac{\partial H}{\partial q_{j}} \right> + \left< \frac{p_{i} p_{j}}{p} \frac{\partial H}{\partial q_{j}} \right> \right) \right. \notag \\
&& \left. - \beta^2 \left( \delta_{ij} \left< p^2 \frac{\partial H}{\partial q_{j}} \right> + 3\left< p_{i} p_{j} \frac{\partial H}{\partial q_{j}} \right> \right) \right\}.
\eer

\noindent Putting these equations into the form
\ber \label{GEEgen1.1}
\frac{d}{dt} \left< q_{i} \right> &=& \sum_{j = 1}^{N} \left\{ \delta_{ij} \left(\left< \frac{\partial H}{\partial p_{j}} \right> - \alpha \left< p \hspace{0.08cm} \frac{\partial H}{\partial p_{j}} \right> + \beta^2 \left< p^2 \frac{\partial H}{\partial p_{j}} \right> \right) \right. \notag \\ 
&& \left. - \alpha \left< \frac{p_{i} p_{j}}{p} \frac{\partial H}{\partial p_{j}} \right> + 3 \beta^2 \left< p_{i} p_{j} \frac{\partial H}{\partial p_{j}} \right> \right\}
\eer
and
\ber \label{GEEgen2.1}
\frac{d}{dt} \left< p_{i} \right> &=& \sum_{j = 1}^{N} \left\{ - \delta_{ij} \left( \left< \frac{\partial H}{\partial q_{j}} \right> - \alpha \left< p \hspace{0.08cm} \frac{\partial H}{\partial q_{j}} \right> + \beta^2 \left< p^2 \frac{\partial H}{\partial q_{j}} \right>  \right) \right. \notag \\
&& \left. + \alpha \left< \frac{p_{i} p_{j}}{p} \frac{\partial H}{\partial q_{j}} \right> - 3 \beta^2 \left< p_{i} p_{j} \frac{\partial H}{\partial q_{j}} \right> \right\}.
\eer
we recognize, in the $\delta{ij}$-term, the pattern $(1 - \alpha \sqrt{p_i p_i} + \beta^2 p_i p_i)$ that arises in various results of the GUP modified angular momentum algebra \cite{our-gup}.

Since, we shall be interested in one dimensional problem of free wave-packet expansion we express the above mentioned equations for one dimension, starting from the commutator,
\be \label{neq}
    \big[ q, p \big]_{GUP} = i \hbar \big( 1 - 2\a p + 4\b^2 p^2 \big),
\ee
which leads to the other commutation relations
\be \no
\big[ q, H \big]_{GUP} = i\hbar \Tilde{\gamma} \frac{p}{m} \hspace{1cm} ; \hspace{1cm} \big[ p, H \big]_{GUP} = - i\hbar \Tilde{\gamma} V',
\ee
where, $\Tilde{\gamma} \equiv 1 - 2\a p + 4 \b^2p^2$. If we consider a free particle, then of course $\big[ p, H \big]$ = 0. Using the above results and \eqref{ident} we find
\be \no
\frac{d}{dt} \left< q \right> = \frac{1}{m} \left< \Tilde{\g} p \right> \hspace{0.5cm}
\ee
and
\be \no
\frac{d}{dt} \left< p^n \right> = \frac{1}{i\hbar} \left< \big[ p^n, H \big] \right> = \frac{1}{i\hbar} \left< \sum_{j = 0}^{n - 1} p^j \big[ p, H \big] p^{n - (j + 1)} \right> = 0
\ee
for the free particle in one dimension. Now we move to the next section to derive the modified master equation with GUP.

\section{One Dimensional Wave Packets and Generalized Master Equation}
\label{gme}
Using the results of \ref{gee} and equation \eqref{ident}, for a free particle we find
\be \label{neqdd}
\ddot \xi_{GUP} = \frac{2}{m^2} \left\{ \left< \Tilde{\g} p^2 \right> - 2\a \left< \Tilde{\g} p^3 \right> + 4\b^2 \left< \Tilde{\g} p^4 \right> - \left< \Tilde{\g} p \right>^2 \right\}.
\ee
As a quick consistency check, let $\a = \b = 0$, i.e., $\Tilde{\g} = 1$; the result is $\ddot \xi = \frac{2}{m^2} \eta_0$, which is just the standard result for a free particle using the HUP. For the future convenience we want to rewrite \eqref{neqdd} in terms of some new variables, in the form 
\be \label{neqdd2}
\ddot \xi_{GUP} = \frac{2}{m^2} \left( \eta_0 - 4\a\Tilde{C_1} + 4\a^2 \Tilde{C_2} + 8\b^2 \Tilde{C_3}\right)
\ee
which is correct upto second order in GUP parameters. This consideration is in line with the very definition  of the GUP  commutator \eqref{GUPgen} which is an approximate expression up to quadratic powers. The new variables in \eqref{neqdd2} are defined as
\ber
\label{cc1} \tilde{C_1} &=& \eta \left( 2p_{cl} + \Gamma_1 \eta^{1/2} \right), \\
\label{cc2} \Tilde{C_2} &=& \eta^2 \left( \Gamma_2 - 1 \right) + 4\eta p_{cl} \left( p_{cl} + \eta^{1/2} \Gamma_1 \right), \\
\label{cc3} \Tilde{C_3} &=& \eta \left( 3p_{cl}^2 + 3\eta^{1/2}p_{cl}\Gamma_1 + \eta\Gamma_2 \right).\eer
where $\eta = \eta_0$ is the square of (constant) standard deviation in momentum which also appeared with HUP. 

Notice that the new variables $\tilde{C_1}$, $\tilde{C_2}$ and $\tilde{C_3}$ involve  \emph{the higher-order moments}, which introduce a novel statistical interpretation to our discussion, regarding the shape of the probability distribution for free wave-packets. To understand this meaningfully, we have   introduced Pearson's \emph{skewness coefficient} ($\Gamma_1$) which represents the third order moment, as
\begin{equation} \label{skewness}
\Gamma_{1} = \frac{\left< \left( p - \left< p \right> \right)^3 \right>}{\sigma^3} = \frac{1}{\eta^{3/2}} \left< \left( p - \left< p \right> \right)^3 \right>.
\end{equation}
Further, we  have also introduced the fourth order moment given by the \emph{kurtosis coefficient} $\Gamma_2$ as
\begin{equation} \label{kurtosis}
\Gamma_{2} = \frac{\left< \left( p - \left< p \right> \right)^4 \right>}{\sigma^4} = \frac{1}{\eta^2} \left< \left( p - \left< p \right> \right)^4 \right>.
\end{equation}
The term $\sigma \equiv \sqrt[]{\left< p^2 \right> - \left< p \right>^2} = \eta^{1/2}$ is the the \emph{standard deviation} of the momentum distribution. It is important to recall that  both $\Gamma_1$ and $\Gamma_2$  measure the departure of the probability distribution from the \emph{normal distribution}. While $\Gamma_1$ measures the {\it asymmetry} about its mean $\left< p \right>$, $\Gamma_2$ measures its \textit{tailedness}. While the skewness can either take positive or negative values, kurtosis is positive definite. For a normal (true Gaussian) distribution $\Gamma_1 = 0$ and $\Gamma_2 = 3$. This is a remarkable result since every new correction coming from the GUP has a distributional interpretation and, therefore, can be explained physically. Just for completeness, let us note the expanded form of $\Gamma$s, given by
\begin{equation*}
\Gamma_1 = \frac{1}{\eta^{3/2}} \left( \left< p^3 \right> + 2\left< p \right>^3 - 3\left< p \right>\left< p^2 \right> \right),
\end{equation*}
and
\begin{equation*}
\Gamma_2 = \frac{1}{\eta^2} \left( \left< p^4 \right> - 4\left< p \right>\left< p^3 \right> + 6\left< p^2 \right>\left< p \right>^2 - 3\left< p \right>^4 \right).
\end{equation*}

\section{GUP modified broadening of wave-packets: The Free Particle}
\label{gup}

It is straightforward, once again, to write down a solution of \eqref{neqdd2}, under the assumption that the initial wavepacket is minimal ($\dot{\xi} (t_0) = 0$). The resulting equation is 
\be \label{gensol}
{\Delta q}_{free}(t) = \left[ \left( \Delta q_0 \right)^2 + \frac{1}{m^2} \left\{ \left( \Delta p_0 \right)^2 - 4\a \tilde{C_1} + 4\a^2\tilde{C_2} + 8\b^2\tilde{C_3} \right\} t^2 \right]^{1/2}
\ee
which depends on $\tilde{C_1}$, $\tilde{C_2}$ and $\tilde{C_3}$ (which carry the information on the standard deviation, skewness and kurtosis of the probability distribution in momentum space), as well as GUP parameters $\a$ and $\b$. From here on, we shall branch our discussion in two directions, with two special cases of the GUP (i) with $\alpha = \beta$, which is the Ali-Das-Vagenas (ADV) form of GUP, and (ii) $\alpha = 0$, which is the Kempf-Mann-Mangano (KMM) form.

\subsection{Ali-Das-Vagenas (ADV) GUP}

We can arrive to the form of GUP prescribed by Ali, Das and Vagenas in \cite{our-gup, our-gup2, thesis}, just by setting $\a = \b$ in the commutator \eqref{neq}, giving us
\begin{equation} \label{gamma}
\big[ q,p \big]_{GUP} = i \hbar \left(1 - 2\alpha p + 4\alpha^2 p^2\right) .
\end{equation}
The solution \eqref{gensol} dictating the spread over time for the free wave packet for this case is given by
\begin{eqnarray}\label{freegup}
\Delta q_{free}(t) = \sqrt{\xi} (t) = \sqrt[]{{\Delta q_{0}}^2 + \frac{1}{m^2} \big( {\Delta p_{0}}^2 - 4\alpha C_{1} + 4\alpha^2 C_{2} \big) t^2 } ,
\end{eqnarray}
where the coefficients $C_1 = \tilde{C_1}$ and $C_{2} = 3\left< p^4 \right> - \left< p^2 \right>^2 - 2p_{cl}\left< p^3 \right>$. 
Now, before going on to the analysis of the GUP-modified spread of free wave-packets, we need to find an expression for $\eta$ as a function of the initial \textit{size} of the wave-packet $\xi_0 = \left( \Delta q_0 \right)^2$. To do this, note that \eqref{GUPgen} leads to the minimum uncertainty relation, which for the ADV form is
\be \label{min-uncert}
\Delta q_0 \Delta p_0 = \frac{\hbar}{2} \left[ 1 + \left( \frac{\a}{\sqrt{\left< p^2 \right>}} + 4\a^2 \right) \Delta p_0^2 + 4\a^2 p_{cl}^2 - 2\alpha \sqrt{\left< p^2\right>} \right] 
\ee
Using this and the fact that $\left< p^2 \right> = \eta_0 + p_{cl}^2$, we find that
\be \label{find-eta}
\frac{2}{\hbar} \left( \Delta q_0 \sqrt{\eta_0} \right) - \left[ 1 + 4\a^2 \left(\eta_0 + p_{cl}^2 \right) \right] + \a \left[ \frac{\eta_0 + 2p_{cl}^2}{\sqrt{\eta_0 + p_{cl}^2}} \right] = 0,
\ee
Upon solving this equation for $\eta_0$ we find the expression $\eta_0 = \eta_0 \left( \D q_0, \alpha, \beta, p_{cl} \right)$ that we were looking for. Notice that, since both $\D q_0$ and $p_{cl}$ are constant parameters that depend on the particle (or molecule) under consideration, and $\b$ is the GUP parameter, solving \eqref{find-eta} will yield a numerical value for $\eta_0$ that will be different for the different systems that one is considering. We shall take advantage of this in the following section.

\subsection{Kempf-Mann-Mangano (KMM) GUP}

The KMM form of GUP (proposed in \cite{Mangano} and further discussed in \cite{g7}) does not include any linear term in the momentum, and is given by
\begin{equation} 
\label{kmm-gamma}
\big[ q,p \big]_{GUP} = i \hbar \left( 1 + {\tilde{\beta}}^2 p^2 \right)
\end{equation}
which is identical to \eqref{neq} with the identification $\tilde{\beta} = 2\beta$. To get the solution for the free wavepacket expansion in this case we can just set $\a = 0$ in \eqref{gensol} and this gives
\be \label{kmm}
{\Delta q}_{free}(t) = \left[ \left( \Delta q_0 \right)^2 + \frac{1}{m^2} \left\{ \left( \Delta p_0 \right)^2 + 2\tilde{\b}^2\tilde{C_3} \right\} t^2 \right]^{1/2}
\ee
and it does not include $\tilde{C_1}$ and $\tilde{C_2}$. However, statistically speaking it has the same interpretation in terms of $\Gamma_1, \Gamma_2, \eta$ since all of them are included  in the definition of $\tilde{C_3}$. The corresponding minimum uncertainty relation and the relationship to find $\eta$ are now given by,
\be \label{kmm-min-uncert}
\Delta q_0 \Delta p_0 = \frac{\hbar}{2} \left( 1 +  4\beta^2  (\Delta p_0^2 + p_{cl}^2) \right). 
\ee
and
\be \label{kmm-find-eta}
\frac{2}{\hbar} \left( \Delta q_0 \sqrt{\eta_0} \right) - \left[ 1 + 4\beta^2 \left(\eta_0 + p_{cl}^2 \right) \right] = 0
\ee
respectively.

\section{Results and Physical Interpretation} \label{results}

Now, let us elaborate on the results obtained in the last sections.

The standard discussion based on the HUP provides a universal time-evolution law \eqref{freeex} for the wave-packet's width, irrespective of the initial probability distribution at time $t_0$. The only requirement for \eqref{freeex} is that the wave-packet's width was minimal at $t_0$. This will apply for a normal distribution (which is quite ideal) and also for all other situations where the initial probability distribution is not normal. For all cases, the evolution law is the same and is given by \eqref{freeex}. On the other hand, as evident from our analysis, that is not true if we have to believe a GUP based calculation, irrespective of the particular form one may choose (such as ADV or KMM form). The modified time evolution laws  \eqref{freegup} and \eqref{kmm} are, indeed, dependent on the type of initial probability distribution. That is to say, for two wave-packets of the same initial width but different form (different value of skewness or kurtosis) the dispersion rate will be different for both  \eqref{freegup} and \eqref{kmm}. With that said, the distributions do not need to be skewed or with excess kurtosis in order to exhibit GUP-induced effects (the evolution of normal Gaussian wave-packets is modified as well). Furthermore, these rates are dependent on \emph{both} the initial momentum and uncertainty in momentum, as opposed to the standard case \eqref{free} where it does not depend on the initial momentum.

One may now ask the question: Why do we have to consider different initial probability distributions, at all, for a free particle? To answer this question we may think about a stream of particles which were under some sort of applied force fields for some time and then those force fields were switched off at time $t_0$, and from that instant on (or a little while after, depending on the relaxation time) these particles start behaving as free wave-packets. Then the initial configuration of the wave-packet at time $t_0$, when all the force fields are switched off, depends on the details of the interaction between the particles and said force fields, which can of course be arbitrary and, therefore, the initial configuration of the stream of free particles at $t_0$ need not be a normal distribution. In fact, it is likely to have any other distribution including the possibility to have a nonzero skewness and excess kurtosis.

Therefore, from our discussion it follows that, while an HUP based calculation is blind to the moments higher than second order of the initial probability distribution in momentum space, GUP based approaches do differentiate between two different initial templates; it shows {\it an enhanced memory of the initial probability distribution (such as skewness and kurtosis) at any later instant of time}. Note that, however, since all of the physical parameters such as the skewness and kurtosis in momentum space, and average momentum are constants in time for a free particle, their initial values will be unchanged during the  course of time. Further, $\eta = \langle p^2 \rangle - \langle p \rangle^2$ is also constant in time for a free particle so that the initial uncertainty in momentum space remains unchanged over the course of time - there is no spreading in momentum space.

\begin{figure}[t!]
\includegraphics[width=0.85\textwidth]{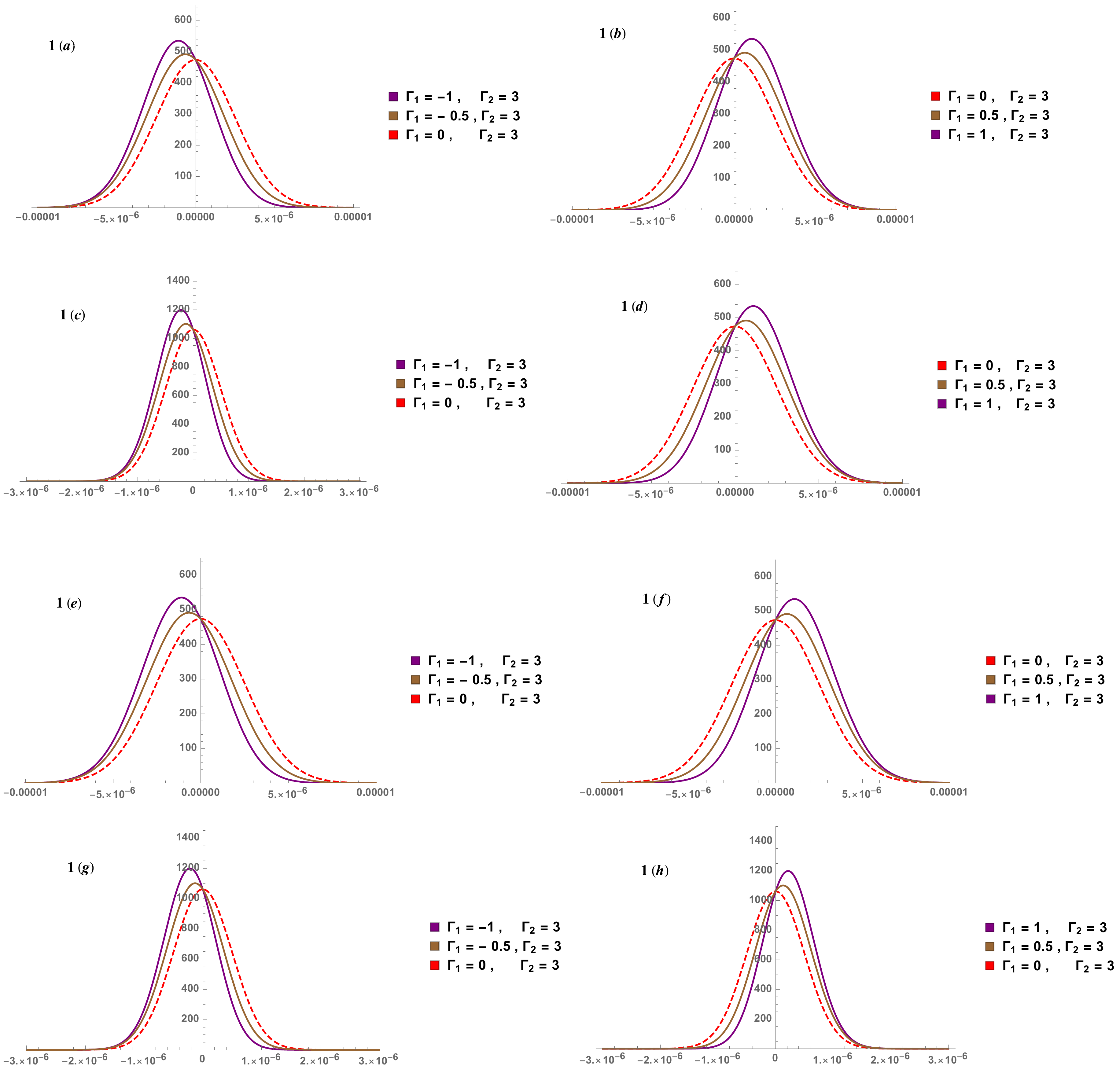}
\caption{Comparison of GUP wavepackets with various skewness and normal kurtosis. The spatial coordinate is along the $X$ axis and the wavefunction is along the $Y$ axis. The left column of figures 1(a), 1(c), 1(e) \& 1(g) correspond to the normal vs negative skewness, whereas, the right column of figures 1(b), 1(d), 1(f) \& 1(h) correspond to the normal vs positive skewness. The KMM form is used in Figs. 1(a) - 1(d) and the ADV form is used in 1(e) - 1(h). The first and third rows 1(a), 1(b), 1(e) \& 1(f) correspond to $C_{60}$, while, the second and fourth rows 1(c), 1(d), 1(g) \& 1(h) correspond to $C_{176}$.}
\label{kmm-var-skew}
\end{figure}

To start analyzing these GUP-induced effects, let us first consider a {\em{skewed}} probability distribution (with vanishing excess kurtosis) of the initial wave-packet. A template of such a wave-packet can be expressed in terms of the following function (we are only considering the positive amplitude)
\be
f({\Gamma_1},{t}) \text{:=} \frac{\exp \left(-\frac{q^2}{2 \xi (t) }\right) \left(\text{erf}\left(\frac{q \Gamma_1}{2 \sqrt{\xi (t)}}\right)+1\right)}{ \left( \pi  \xi (t) \right)^{1/4}}.
\label{skewf}
\ee
It is easy to check that the probability distribution associated to this wave-packet (that is, the square of \eqref{skewf}) is normalized over the configuration space and, therefore, satisfies the probability conservation condition at all times. This function corresponds to a skewed distribution with normal kurtosis $\Gamma_2 = 3$; its width satisfy, depending on the form of GUP, the equations  \eqref{kmm} or \eqref{freegup}, and for a given instant of time $t$ the shape of the wave-packet will change for a given value of the skewness $\Gamma_1$. Fig. \ref{kmm-var-skew} we plot this behavior for both $C_{60}$ and $C_{176}$ ``buckyball" molecules. The details of parameter values are given in the figure. We chose $C_{60}$ and $C_{176}$, also known as the ``buckyballs" (scientific name \textit{Buckminsterfullerene}), molecules for this analysis because they are one of the, commercially available, bigger-sized molecules that behave as a single wave-packet, thus they can used for experimental studies on our proposal. This point will be further clarified in the next section where we discuss a possible test of our results. Just for the reminder, plots with positive skewness have more probability that the particle will be found on the right side than the left side of the mean value and vice-versa. Here we have assumed characteristic values for several parameters including the mass and the initial \textit{size} (taken to be the van der Waals diameter \cite{buckysize}) of the molecule. Note that the GUP coupling constants $\alpha$ and $\beta$ are taken to be order 1. These plots are therefore more for a qualitative understanding. Accurate quantitative analysis for testing our result will be carried out in the next section.

Notice that, even though $\Gamma_1$ and $\Gamma_2$ are defined as the skewness and kurtosis coefficients in {\em momentum space} (see \eqref{skewness} and \eqref{kurtosis}), this does of course introduce skewness and kurtosis in position space as well, so that the shape of the wave-packet in position space will also be affected, as shown in the figures. The difference is that the skewness and kurtosis coefficients in {\em position space} will change over time; we can readily see this from the fact that, generally speaking, $\left< q^n \right> = \left< q^n \right> (t)$ for the free particle. With that said, notice that the GUP-modified spread evolution laws for free wave-packets \eqref{kmm} and \eqref{freegup} do {\em not} depend explicitly on these coefficients in position space, but rather in momentum space, so we do not need to compute the former for our present analysis.

\begin{figure}[t!]
\includegraphics[width=0.85\textwidth]{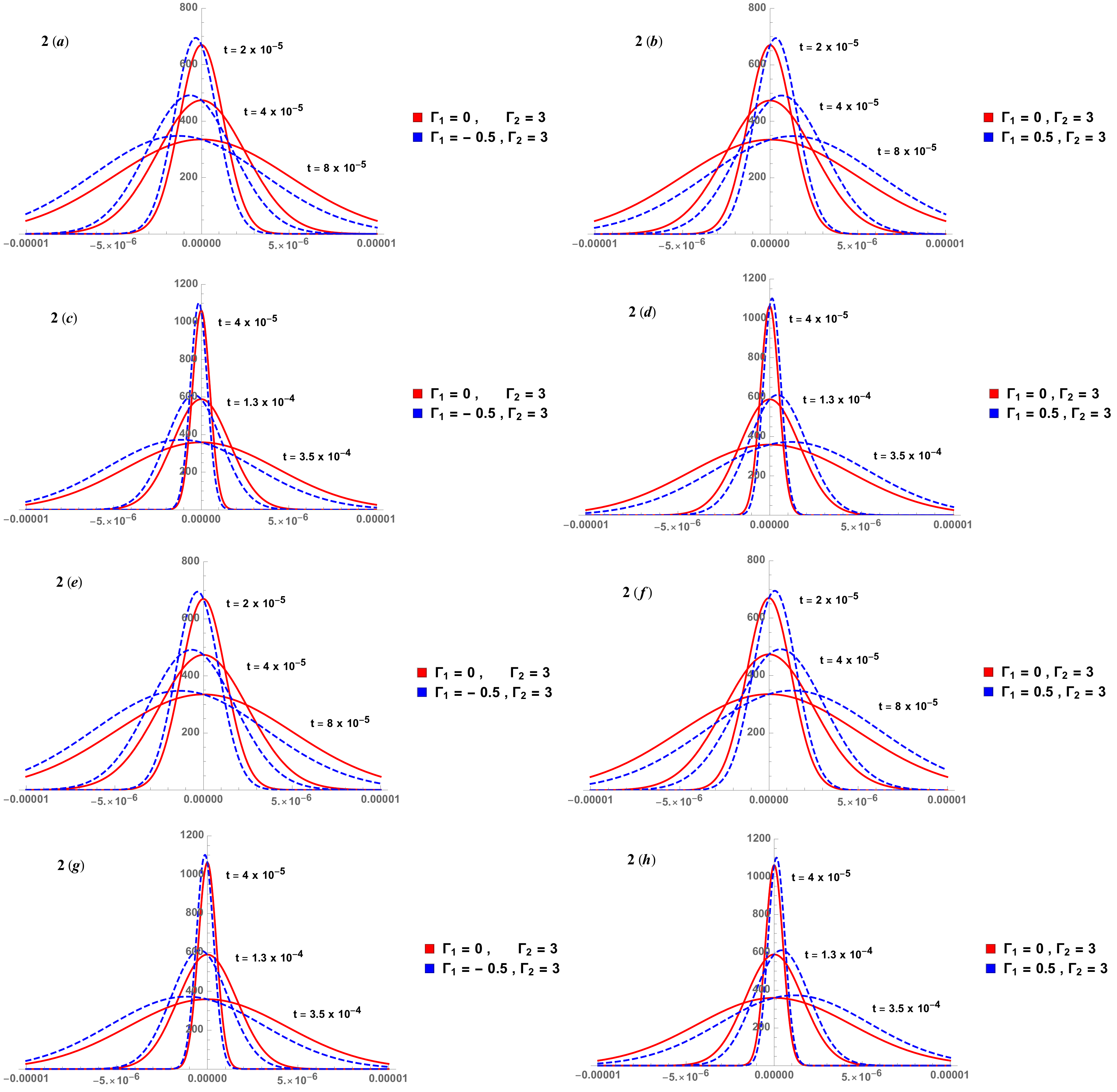}
\caption{Comparison of (all) GUP time evolution  between the normal vs skewed wave-packets for both KMM and ADV forms. The spatial coordinate is along the $X$ axis and the wavefunction is along the $Y$ axis. The left column of figures 2(a), 2(c), 2(e) \& 2(g) correspond to the negative skewness, whereas, the right column of figures 2(b), 2(d), 2(f) \& 2(h) correspond to the positive skewness. The KMM form is used in Figs. 2(a) - 2(d) and the ADV form is used in 2(e) - 2(h). The first and third rows 2(a), 2(b), 2(e) \& 2(f) correspond to $C_{60}$, while, the second and fourth rows 2(c), 2(d), 2(g) \& 2(h) correspond to $C_{176}$.}
\label{skew-time}
\end{figure} 

Now, let us plot the time evolutions of this wave-packet governed by \eqref{skewf}, in Fig. \ref{skew-time}, for both $C_{60}$ and $C_{176}$ parameters for both KMM and ADV time evolutions \eqref{kmm} and \eqref{freegup}. Again, the shape and the rate at which it spreads depends on the value of $\Gamma_1$ appearing in \eqref{kmm} and \eqref{freegup} (through $\tilde{C_3}$ in \eqref{kmm} and through both $C_1$and $C_2$ in \eqref{freegup}). Clearly, the initial distribution has an important role to play in the time evolution of the wave-packet, and this is a new insight coming from the GUP based analysis, again irrespective of the KMM or ADV  forms that one may consider - both modifications have the same statistical interpretation.

\begin{figure}[t!]
\includegraphics[width=0.85\textwidth]{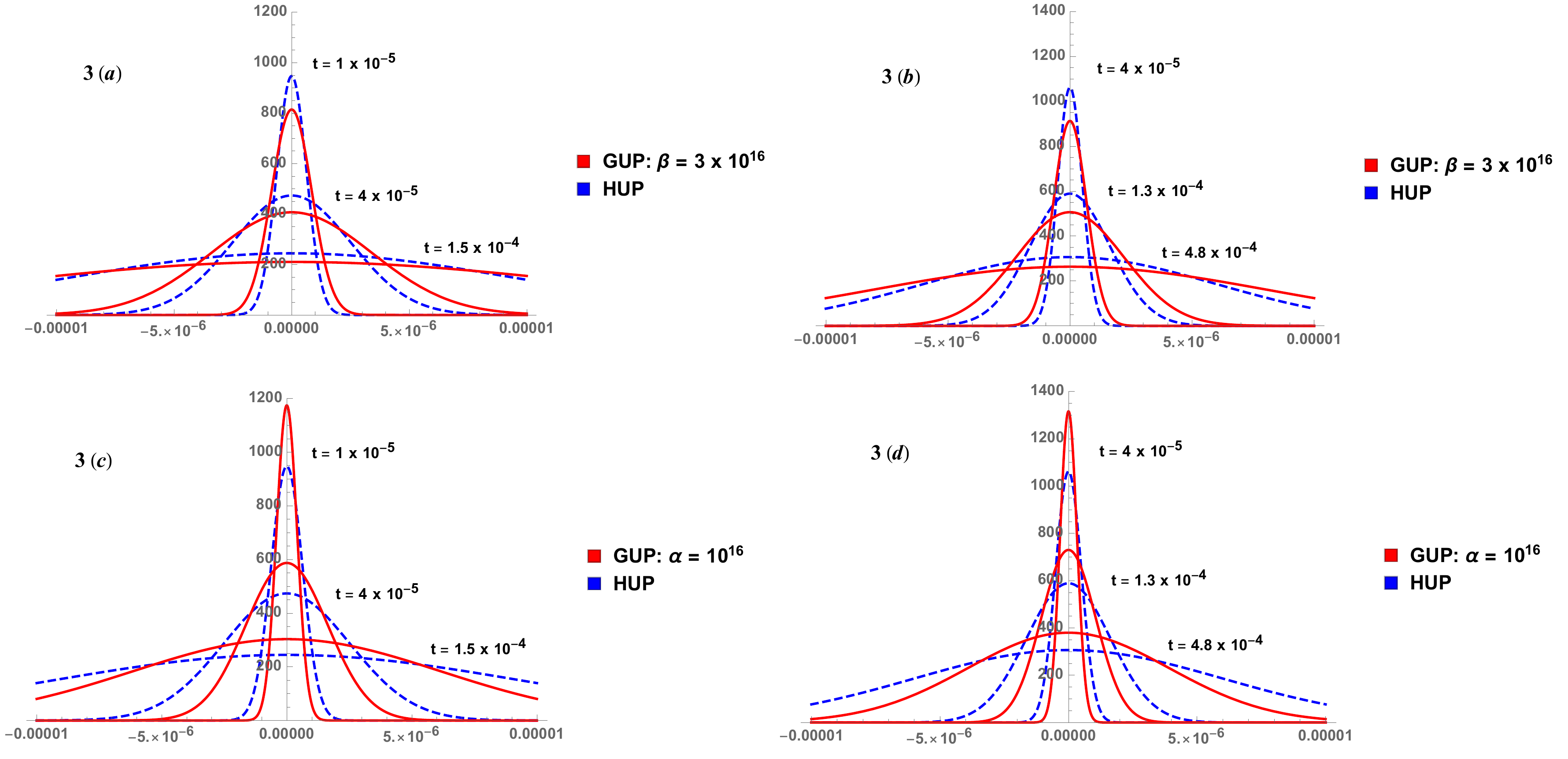}
\caption{GUP vs HUP time evolution (broadening) of free wave-packets. The spatial coordinate is along the $X$ axis and the wavefunction is along the $Y$ axis. The left column of figures 3(a) and 3(c) are for $C_{60}$ and the right column 3(b) and 3(d) for $C_{176}$ molecule. The first row is for KMM and the second row is for ADV form. For more discussion see text.}
\label{GvsH}
\end{figure}

In figure  \ref{GvsH}, we compare the wave-packet evolution with and without the GUP modifications. The sample distribution is again given by \eqref{skewf} with either $C_{60}$ or $C_{176}$ parameters, and we consider the normal (Gaussian) part of it by setting $\Gamma_1 = 0$. We find some important insights by looking at these plots: first, for the KMM GUP \eqref{kmm-gamma} the minimum uncertainty wave-packet, defined at the initial time, has a larger width for the GUP-based calculation than the HUP-based standard result, whereas, for the ADV GUP \eqref{gamma} it is opposite - the minimal wavepacket has a smaller width than the HUP based minimal width. It is therefore consistent to say that for the ADV form of GUP, for a physical quantum system, such as the one given by these ``buckyballs'', the existence of a minimal length scale in the form of \eqref{gamma} minimizes the uncertainty in the probability distribution in position space for the same momentum distribution. Interestingly, this ADV result of further squeezing the free, minimal wavepacket, may be related, of course with certain differences, with an expectation that gravity might have a natural tendency to localize the wavefunction, as first pointed out by Penrose and D\'iosi \cite{penrose2}. This localization process could be evident in ADV form. However, the KMM form predicts the opposite behavior where the minimal wavepacket increases its width as compared to the standard HUP case, due to their specific form of GUP. The interpretation of this behavior is not well known. 

Furthermore, as we shall see in the next section, the broadening rate for the KMM form will be quicker than the HUP broadening rate, whereas, the ADV form will predict a slower broadening rate for the minimal width wavepacket, for a vast range of parameter values. 

\begin{figure}[t!]
\includegraphics[width=0.7\textwidth]{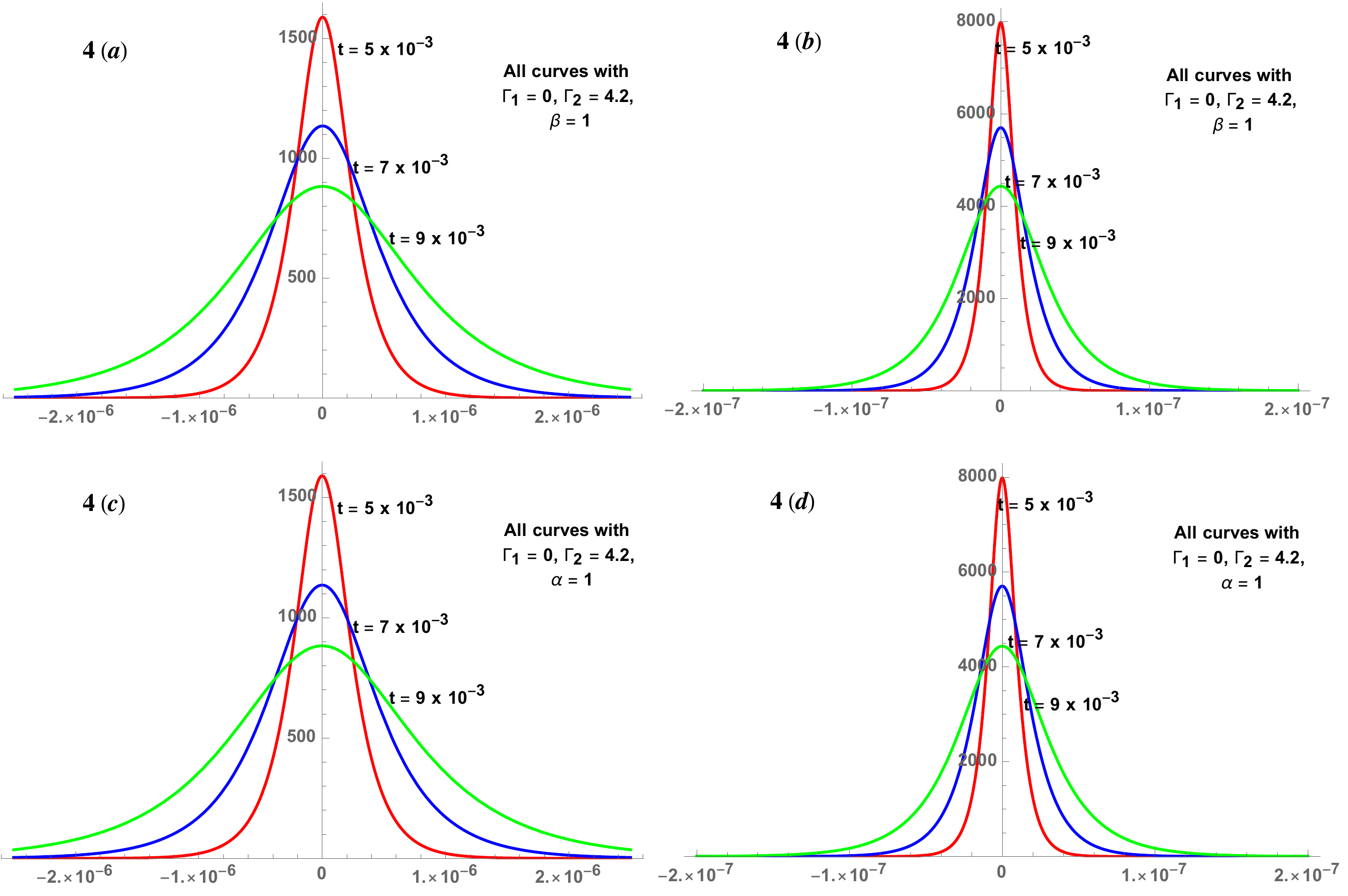}
\caption{GUP modified free wave-packets with fixed excess kurtosis. The spatial coordinate is along the $X$ axis and the wavefunction is along the $Y$ axis. The left column of figures 4(a) and 4(c) are for $C_{60}$ and the right column 4(b) and 4(d) are for $C_{176}$ molecule. The first row is for KMM and the second row is for ADV form.}
\label{nor-kurt}
\end{figure}

\begin{figure}[t!]
\includegraphics[width=\textwidth]{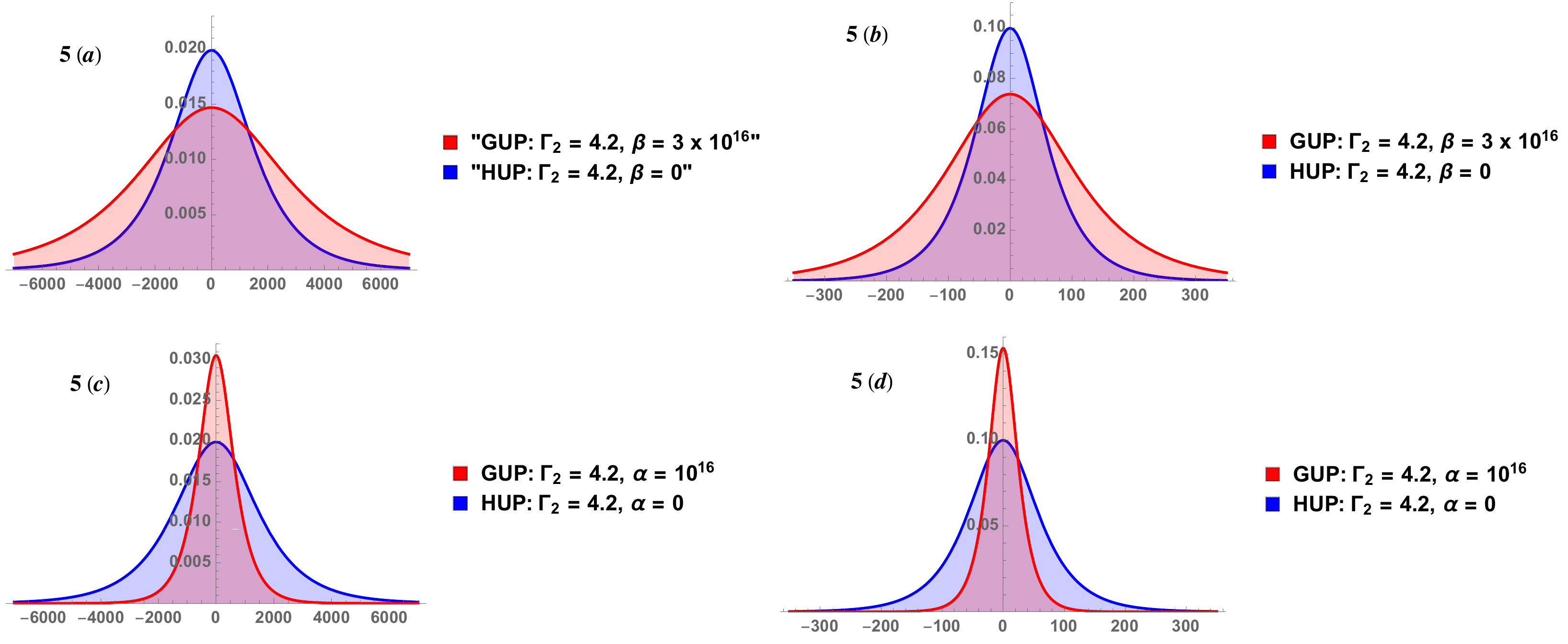}
\caption{Comparison between wave-packets predicted by GUP and HUP for a probability distribution with excess kurtosis. The spatial coordinate is along the $X$ axis and the wavefunction is along the $Y$ axis. The left column of figures 5(a) and 5(c) are for $C_{60}$ and the right column 5(b) and 5(d) are for $C_{176}$ molecule. The first row is for KMM and the second row is for ADV form.}
\label{gvs-nor-kurt}
\end{figure}

So far our discussion did not include a distribution with an excess kurtosis. In order to study this let us assume the probability density function of the logistic distribution, given by 
\be
G(q,\xi (t))\text{:=}\frac{\exp \left(-\frac{q}{\xi (t)}\right)}{\xi(t) \left(\exp \left(-\frac{q}{\xi(t)}\right)+1\right)^2}.
\label{gq}
\ee
This function has skewness $\Gamma_1 = 0$ and excess kurtosis with $\Gamma_2 = 4.2$. The wave-packet associated with this probability distribution is (again considering the positive amplitude)
\be \label{kurt-wp}
g \left(q, \xi(t) \right) \text{:=} \frac{\exp \left(-\frac{q}{2 \xi (t)}\right)}{ \sqrt{\xi(t)} \left(\exp \left(-\frac{q}{\xi(t)}\right)+1\right)} .
\ee
The rate at which the spreading takes place with the GUP modifications, both for the KMM \eqref{kmm} and ADV \eqref{freegup} forms for \eqref{kurt-wp} include the kurtosis $\Gamma_2$ as opposed to the standard prediction from the HUP, where the rate of expansion of $\xi(t)$ \eqref{free} is independent of the value of kurtosis. In Fig. \ref{nor-kurt} we plot the GUP time evolution of \eqref{kurt-wp} starting from the minimal width wave-packet for KMM and ADV forms of GUP. Figures 4(a) and 4(b) belong to the KMM form and correspond to (a) $C_{60}$ and (b) $C_{176}$ molecules. Whereas, figures 4(c) and 4(d) belong to the ADV form and correspond to (c) $C_{60}$ and (d) $C_{176}$ molecules. Note that, when considering any initial distribution (be it normal, skewed or with excess kurtosis), if one takes the GUP parameters $\alpha, \tilde{\beta} \sim 1$, the time evolution is practically identical to the HUP based calculation and it is hard to differentiate between the two in the plots irrespective of KMM or ADV forms. However, given that the allowed parameter space for $\tilde{\beta},~\alpha$ is quite wide \cite{bounds}, for larger values of these parameters these plots do show a significant difference between the width of the wave-packet with or without GUP corrections. This can be visualized in Fig. \ref{gvs-nor-kurt} where we have considered values $\a,~\tilde{\b} \sim {\cal O} (10^{16})$ for both KMM and ADV GUP parameters. Figures, 5(a) and 5(b) belong to the KMM form while 5(c) and 5(d) belong to the ADV form. This characteristic of time evolution and its difference with or without GUP modifications is just similar as before that we considered in Fig. \ref{GvsH} - just that here we plot one snapshot while in \ref{GvsH} we have three of them. In the next section we shall speak more about the numbers and the likelihood of measuring them in realistic experiments.

\section{Possible tests}
\label{tests}

In this section we study the possibility of experimental verification of the minimal length effect on the dispersion of the free wave-packets. The scheme that we propose here is quite simple -  {\em one needs to measure the timescale in which the wave-packet (describing a particle or a system of particles behaving as a single wave-packet) doubles its initial width}. In fact, one can choose any final size that is permissible, but our calculation here will be done considering that the wave-packet is doubling its size.  HUP based calculation gives a precise estimate for that which we already discussed for the case of electrons in section II. 

Let us re-do the analysis, now in presence of the GUP modifications. Clearly the doubling time will be different depending on the choice of modifying the commutator such as KMM or ADV definitions. For the KMM case, using \eqref{kmm} this doubling time is found to be
\be\label{tgup-kmm}
t_{\text{double}}^{\text{KMM}} = \frac{\sqrt{3} m \Delta q_0}{\sqrt{\Delta p_0^2 + 2\tilde{\beta}^2 \tilde{C_3}}},
\ee
whereas, for the ADV case using \eqref{freegup} we can easily calculate this time to be
\be
t_{\text{double}}^{\text{ADV}} = \frac{\sqrt{3} m \Delta q_0}{\sqrt{\Delta p_0^2 - 4\alpha C_1 + 4\alpha^2 C_2}}
\label{tgup}
\ee
where the minimum uncertainty wave-packet now satisfies either  \eqref{kmm-min-uncert} or \eqref{min-uncert}, depending on the form of GUP under consideration. Plugging in our expressions for $C_1$, $C_2$ and $\tilde{C_3}$, we get
\be \label{kmm-tgup2}
t_{\text{double}}^{\text{KMM}} = \frac{\sqrt{3} m \Delta q_0}{\sqrt{\Delta p_0^2 + 2 \tilde{\beta}^2 \eta \left[ 3 p_{cl} \left( p_{cl} + \eta^{1/2} \Gamma_1 \right) + \eta \Gamma_2 \right] }}
\ee

\be \label{tgup2}
t_{\text{double}}^{\text{ADV}} = \frac{\sqrt{3} m \Delta q_0}{\sqrt{\Delta p_0^2 + 4 \eta \big[ \alpha^2 \left( \left( 3\Gamma_2 - 1 \right) \eta + 10 p_{cl} \left( \Gamma_1 \eta^{1/2} + p_{cl} \right) \right) - \alpha \left( 2 p_{cl} + \Gamma_1 \eta^{1/2} \right) \big]}}
\ee
If we, for the sake of simplicity, consider a Gaussian wave-packet, then we can set the skewness and kurtosis coefficients to $\Gamma_1 = 0$ and $\Gamma_2 = 3$, respectively. With this the above expressions get simplified, giving
\be \label{kmm-tgup-normal}
t_{\text{double}}^{\text{KMM}} (\Gamma_1 = 0, ~\Gamma_2 = 3) = \frac{\sqrt{3} m \Delta q_0}{\sqrt{\Delta p_0^2 + 6 \tilde{\beta^2} \eta \left( p_{cl}^2 + \eta \right)}}
\ee
and
\be \label{tgup-normal}
t_{\text{double}}^{\text{ADV}} (\Gamma_1 = 0, ~\Gamma_2 = 3) = \frac{\sqrt{3} m \Delta q_0}{\sqrt{\Delta p_0^2 + 8 \eta \big( \alpha^2 \left( 4\eta + 5 p_{cl}^2 \right) - \alpha p_{cl} \big)} }
\ee

With expressions \eqref{kmm-tgup-normal} and \eqref{tgup-normal} at hand, we can use the relations \eqref{kmm-find-eta} and \eqref{find-eta} to replace $\eta = \Delta p_0 ^2$ in terms of $\Delta q_0$ and other parameters. Therefore, we now have everything we need for doing a numerical calculation with realistic molecular wave-packets.

First, let us go back to the case of the free electron (where the initial wave-packet had a width of $10^{-10}~m$)  so we can use \eqref{kmm-tgup-normal} and \eqref{tgup-normal} to estimate the magnitude of the GUP modification for both the KMM and ADV forms, respectively. 

For the KMM form, a simple numerical check for the free electron case shows that, for the values $1 \leq \tilde{\beta} \leq 10^{16}$ the difference between the HUP and KMM-GUP predictions is negligible. If we go to values like $\tilde{\beta} = 10^{17}$ we get a difference between both predictions ${\cal O} (10^{-30}) ~s$ -- not anywhere near a potentially detectable value. Going to larger values like $\tilde{\beta} = 10^{22}$ we get a difference ${\cal O} (10^{-20}) ~s$ and even larger values like $\tilde{\beta} = 10^{30}$ give a time difference of $4.06218 \times 10^{-16} ~s$ which is somewhat close to a potentially detectable value.

The numbers are better for the ADV form for the free electron case, but the results are still effectively the same with or without GUP in the parameter range $1 \leq \a \leq 10^{21}$. The numerical calculations show that the difference between HUP and ADV-GUP is at most ${\cal O} (10^{-30}) ~s$ for $\a \leq 10^{10}$, and for higher values like $\a = 10^{17}$ we start getting differences ${\cal O} (10^{-23}) ~s$, but practically undetectable still.

It is therefore clear that for both KMM and ADV forms a free electron wavepacket expansion is almost identical to the original HUP results -- the difference being unlikely to be detected even with the utmost precise of atomic clocks available today. Furthermore, if we have to believe an upper bound for $\alpha$  or $\tilde{\beta}$ we can infer that the GUP modification does not give a major difference in the doubling time for the case of free electrons, at least in the initial stage where it is more likely to be detectable by a laboratory based experiment.

In order for these effects to be detectable in a laboratory, we must magnify the GUP modifications somehow. To do this, we must consider probes whose wave-packets have initial size and associated mass {\em bigger} than that of an electron wavepacket. One  obvious way to achieve this is to consider atoms instead of electrons or, even better, use bigger molecules which can behave like a single wave-packet. This brings us to the so called ``buckyball'' systems and Large Organic Molecules (LOM). ``Buckyballs'' or {\em Buckminsterfullerene} molecules are basically a bunch of carbon atoms behaving as a single quantum wave-packet \cite{c60}.  We shall consider again $C_{60}$ and $C_{176}$ molecules - which we already considered in various plots in the last section. On the other hand, LOMs are probably the most exciting candidates since they are the largest molecules (in terms of the combination of size and mass scale) found so far which behave like a single wave-packet \cite{LOM}.\\

Below we do an analysis for these three objects, using both forms, where we shall keep the relevant GUP parameter  $\alpha$ or $\tilde{\beta}$ as a free parameter from the beginning, and see how the wide range of values for these parameters affect the time difference between the HUP and GUP results for the minimal width wave-packet to double its initial width.

\subsection{Ali-Das-Vagenas (ADV) GUP}
In the case of a $C_{60}$ buckyball molecule, with a mass of $1.19668 \times 10^{-24}$ kg (720 u) and an initial width $\Delta q_0$ equal to its van der Waals diameter (7 \AA) \cite{buckysize}, the HUP prediction for the \textit{doubling} time is $t_{\text{double}} (C_{60}, ~\text{HUP}) = 1.92719 \times 10^{-8} s$. If we start considering ADV type GUP modifications, first with $\a = 1$ as the value of the GUP parameter, then we get practically the same value $t_{\text{double}}$;  the difference between them being 
\be \no
t_{\text{double}}^{\text{ADV}} (C_{60}, ~\a = 1) - t_{\text{double}} (C_{60}, ~\text{HUP})  = - 6.61744 \times 10^{-24} s
\ee
However, if we take large values like $\a = 10^{10}$, then we find that
\be \no
t_{\text{double}}^{\text{ADV}}(C_{60}, ~\a = 10^{10}) - t_{\text{double}} (C_{60}, ~\text{HUP}) = 1.15631 \times 10^{-14} s,
\ee
and if we go even further, like $\a = 10^{16}$, we find $t_{\text{double}} (C_{60}, ~\a = 10^{16}) = 2.96189 \times 10^{-8} s$, and
\be \no
t_{\text{double}}^{\text{ADV}} (C_{60}, ~\a = 10^{16}) - t_{\text{double}} (C_{60}, ~\text{HUP}) = 1.0347 \times 10^{-8} s.
\ee
That is, the difference between both predictions is of the order of the original HUP prediction ($\sim 10^{-8} s$) while taking $\a \sim 10^{16}$ as the GUP parameter.

This analysis shows that depending on the wide range of values for $\a$, the difference between the HUP and GUP predictions for $t_{\text{double}}$ for $C_{60}$ buckyballs stays in an interval where the lower end is undetectable even with the most precise clocks currently available, but the upper end stays well within the available range of precision. 

Furthermore, since we want to amplify the GUP-induced effects (and thus make them easier to detect at laboratory-based experiments), let us now consider a $C_{176}$ buckyball. Using this molecule's parameters ($m = 3.50706 \times 10^{-24}$ kg (2112 u) and $\Delta q_0 = 1.2$ nm \cite{buckysize}), we find that the HUP prediction for the doubling time is $t_{\text{double}} (C_{176}, ~\text{HUP}) = 1.6598 \times 10^{-7} s$ and, again, taking small values of $\a$ (like order unity) yields an effectively undetectable difference between the HUP and GUP predictions. However, if we again set $\a = 10^{10}$, we get
\be \no
t_{\text{double}}^{\text{ADV}} (C_{176},  ~\a = 10^{10}) - t_{\text{double}} (C_{176}, ~\text{HUP}) = 9.9588 \times 10^{-14} s,
\ee
which is better by a factor more than 8 as compared with $C_{60}$, and going to higher values like $\a = 10^{16}$ yields $t_{\text{double}} (C_{176}, ~\a = 10^{16}) = 2.55094 \times 10^{-7} s$ and
\be \no
t_{\text{double}}^{\text{ADV}} (C_{176}, ~\a = 10^{16}) - t_{\text{double}} (C_{176}, ~\text{HUP}) = 8.9114 \times 10^{-8} s.
\ee
This is again an improvement by a factor of almost 9 over the time difference ($1.0347 \times 10^{-8} s$) that we got for the $C_{60}$ molecule.  Therefore, we see that bigger (larger van der Waals diameter) and more massive molecules tend to show stronger deviations from the HUP behavior when considering GUP-modified $t_{\text{double}}$ calculations.

Now let us consider the case of recently discovered LOM wave-packets \cite{LOM}. Considering a TPPF152 or {\em tetraphenylporphyrin} molecule (which consists of 430 atoms and is formally known as $C_{168}H_{94}F_{152}O_{8}N_{4}S_{4}$), with a mass of $5,310$ u ($\sim 8.81746 \times 10^{-24}$ kg) and an initial \textit{size} of 60 \AA.  Taking $\a = 1$, once again, does not bring the time difference in a detectable range. However, if we go to larger values of $\a$ like $10^{10}$ we find
\be \no
t_{\text{double}}^{\text{ADV}} (\text{TPPF152}, ~\a = 10^{10}) - t_{\text{double}} (\text{TPPF152}, ~\text{HUP})  = 6.25961 \times 10^{-12} s.
\ee
which improves the result of the $C_{176}$ molecule by a factor of 63 and this number is 500 times better than for the $C_{60}$ molecule. Further, moving to $\a \sim 10^{16}$ the difference becomes
\be \no
t_{\text{double}}^{\text{ADV}} (\text{TPPF152}, ~\a = 10^{16})  - t_{\text{double}} (\text{TPPF152}, ~\text{HUP}) = 5.60129 \times 10^{-6} s.
\ee 
which is again better by a factor of 63 from $C_{176}$ and 560 from $C_{60}$.

\begin{figure}[t!]
\includegraphics[width=0.9\textwidth]{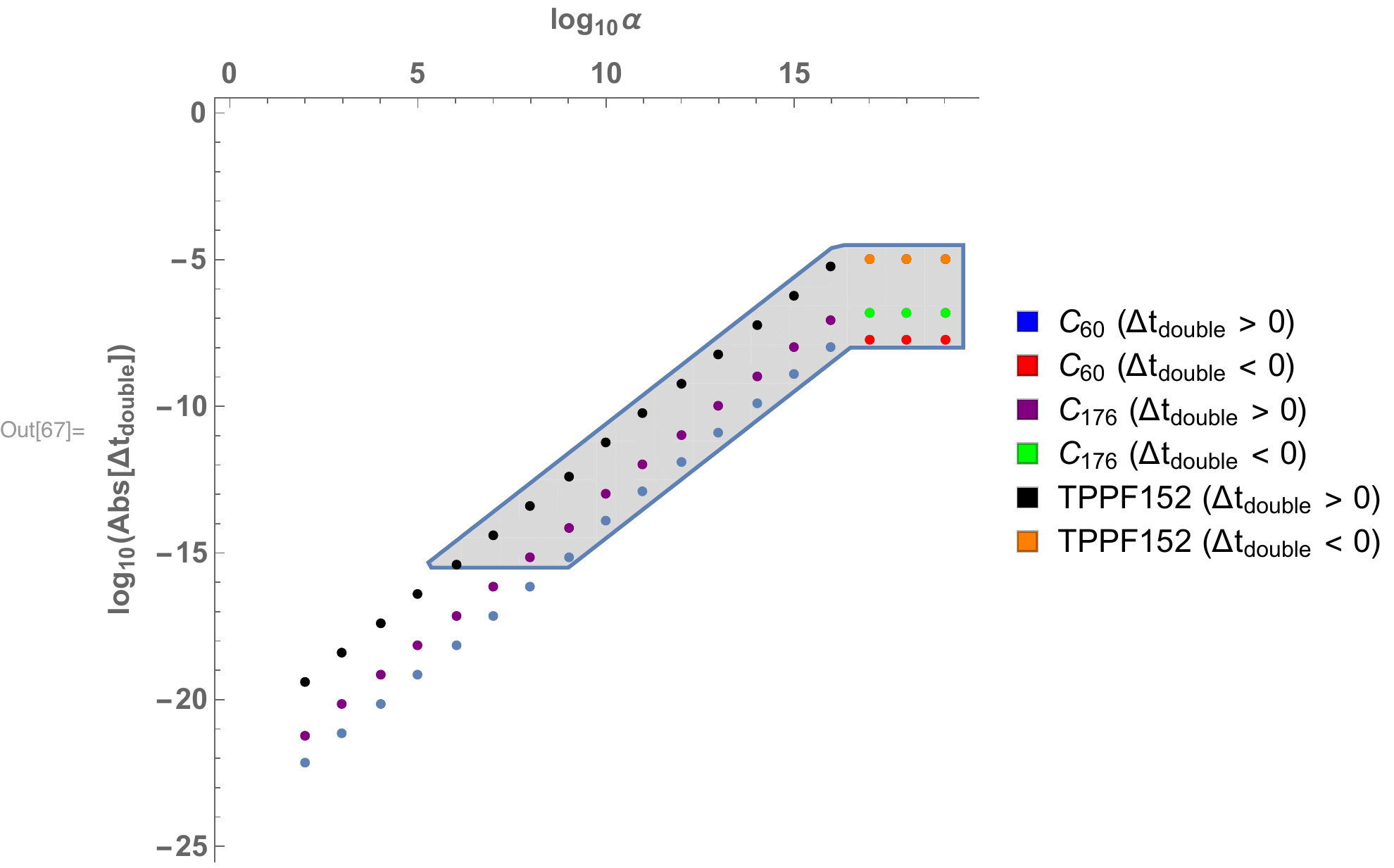}
\caption{$\log$ - $\log$ plots between the GUP parameter $\alpha$ and the doubling time difference $\Delta t_{\text{double}}^{\text{ADV}} = t_{\text{double}}^{\text{ADV}} (\text{GUP}) - t_{\text{double}} (\text{HUP})$ between the GUP and HUP time evolution for (a) $C_{60}$ molecule (the lower plot), (b) $C_{176}$ molecule (the middle plot) and (c) the large organic molecule TPPF152 (the upper plot). The $X$ axis represents the GUP parameter and the $Y$ axis represents the difference between GUP and HUP doubling time. The shaded region indicates the region of parameter space that can be probed by the above molecular wave-packets with an atomic clock of maximum precision $10^{-15}~s$.}
\label{llplot}
\end{figure}

In Fig. \ref{llplot}, we have plotted the difference between the doubling times for various values of $\alpha$ (difference between the ADV type GUP-based and HUP based calculations). This is a log-log plot where values of $\Delta t_{\text{double}}^{\text{ADV}}$ are shown for the parameter space $1\le \alpha \le 10^{19}$. Note that for the larger values of $\alpha \ge 10^{16}$ we get a doubling time difference ${\cal O} (10^{-8}~s)$ for $C_{60}$ molecule which can be easily detected by today's atomic clocks. This result is even better ($10^{-7}~s$) for $C_{176}$ and in $\mu s$ range for TPPF152. On the other side, we can scan the complete parameter space of $\alpha$ (up to order unity), if we can measure a time difference of the order of $10^{-21}~s$ to $10^{-23}~s$, just by considering these molecules. However, if we have to believe that highly precise atomic clocks can differentiate the time measurement by at most $10^{-15}~s$, the use of $C_{60}$ molecules can scan the parameter space $\alpha  \ge10^{9}$, and it is again better for $C_{176}$, for which we can scan $ \alpha \ge10^{8}$. The best of the three, however, stands for TPPF152 which can scan, on the lower side, down to $ \alpha \sim 10^{6}$.  Therefore, if the experiments with TPPF152 do not show the deviation in doubling time within femto-second  we automatically get an improvement by four orders of magnitude on the best existing bound found coming from the Lamb shift \cite{stb}, which is $\alpha \le 10^{10}$. In addition, if we are lucky and Nature behaves in such a manner, we might be able to verify \eqref{gamma} with these molecular wave-packets. If not, we can put a new bound and move on to redo the experiments with even bigger and more massive wave-packets, which could scan the whole parameter space. This is a totally new avenue, that has not been proposed before. {\it In fact any departure from HUP, irrespective of the manner it differs, will be a pathbreaking discovery since it will anyway challenge the standard quantum mechanical prediction}. We expect, perhaps colleagues from the experimental side will find this result interesting.

\subsection{Kempf-Mann-Mangano (KMM) GUP}

Performing a similar analysis for the KMM form \eqref{kmm-gamma} we find that, for the case of the spreading wavepacket of a $C_{60}$ molecule, for values of the GUP parameter $1 \leq \tilde{\b} \leq 10^{9}$ the difference between the HUP-predicted and GUP-predicted {\em doubling} times is practically negligible and far from being in a detectable range. Even going to values $\tilde{\b} \sim 10^{10}$ still gives a very small value for this difference
\be \no
t_{\text{double}}^{\text{KMM}} (C_{60}, ~\tilde{\b} = 10^{10}) - t_{\text{double}}(C_{60}, ~\text{HUP})= - 7.74241 \times 10^{-22} s,
\ee
which is unlikely to be detected even with today's best atomic clocks.

Now, going upto values $\sim {\cal O} (10^{16})$ of $\tilde{\b}$ yields a much more optimistic result
\be \no
t_{\text{double}}^{\text{KMM}} (C_{60}, ~\tilde{\b} = 10^{16}) - t_{\text{double}}(C_{60}, ~\text{HUP})= - 7.38707 \times 10^{-10} s,
\ee
which might be well within the range of precision of the current technology.

If we start to consider $C_{176}$ molecules, again $1 \leq \tilde{\b} \leq 10^9$ yields practically negligible results, and $\tilde{\b} \sim {\cal O} (10^{10})$ gives
\be \no
t_{\text{double}}^{\text{KMM}} (C_{176}, ~\tilde{\b} = 10^{10}) - t_{\text{double}}(C_{176}, ~\text{HUP})= - 6.64391 \times 10^{-21} s,
\ee
an improvement of almost one order of magnitude over the $C_{60}$ result. On the other hand, $\tilde{\b} \sim {\cal O} (10^{16})$ gives
\be \no
t_{\text{double}}^{\text{KMM}} (C_{176}, ~\tilde{\b} = 10^{16}) - t_{\text{double}}(C_{176}, ~\text{HUP})= - 6.36215 \times 10^{-9} s
\ee
which is, again, a much better result. 

Notice that, up to this point, all values of $t_{\text{double}}(\text{GUP}) - t_{\text{double}}(\text{HUP})$ have been negative for the KMM form -- meaning that the KMM-predicted spreading is faster than predicted by the HUP, as opposed to the ADV-based predictions (at least for a majority range of values of the GUP parameter).

Finally, taking our analysis again to the TPPF152 large organic molecule gives much better values of the {\em doubling} time difference, where taking $\tilde{\b} \sim {\cal O} (10^{10})$ gives
\be \no
t_{\text{double}}^{\text{KMM}} (\text{TPPF152}, ~\tilde{\b} = 10^{10}) - t_{\text{double}}(\text{TPPF152}, ~\text{HUP}) = - 4.1674 \times 10^{-19} s.
\ee
Going even further to $\tilde{\b} \sim {\cal O} (10^{16})$ yields
\be \no
t_{\text{double}}^{\text{KMM}} (\text{TPPF152}, ~\tilde{\b} = 10^{16}) - t_{\text{double}}(\text{TPPF152}, ~\text{HUP}) = - 3.99893 \times 10^{-7} s
\ee
which, as is the case for Large Organic Molecules in both types of GUP, is well within the range of possible detection in experiments and once again gives results $\sim 63$ times better than the $C_{176}$ case and $\sim 500$ times better than for $C_{60}$ molecules -- albeit the time differences predicted by a KMM form of GUP are considerably smaller than those predicted by an ADV-type modification of the HUP. This is because the deformation of the commutator only includes a term of quadratic order in momentum and not a term of linear order in momentum in the KMM case.

\begin{figure}[t!]
\includegraphics[width=0.7\textwidth]{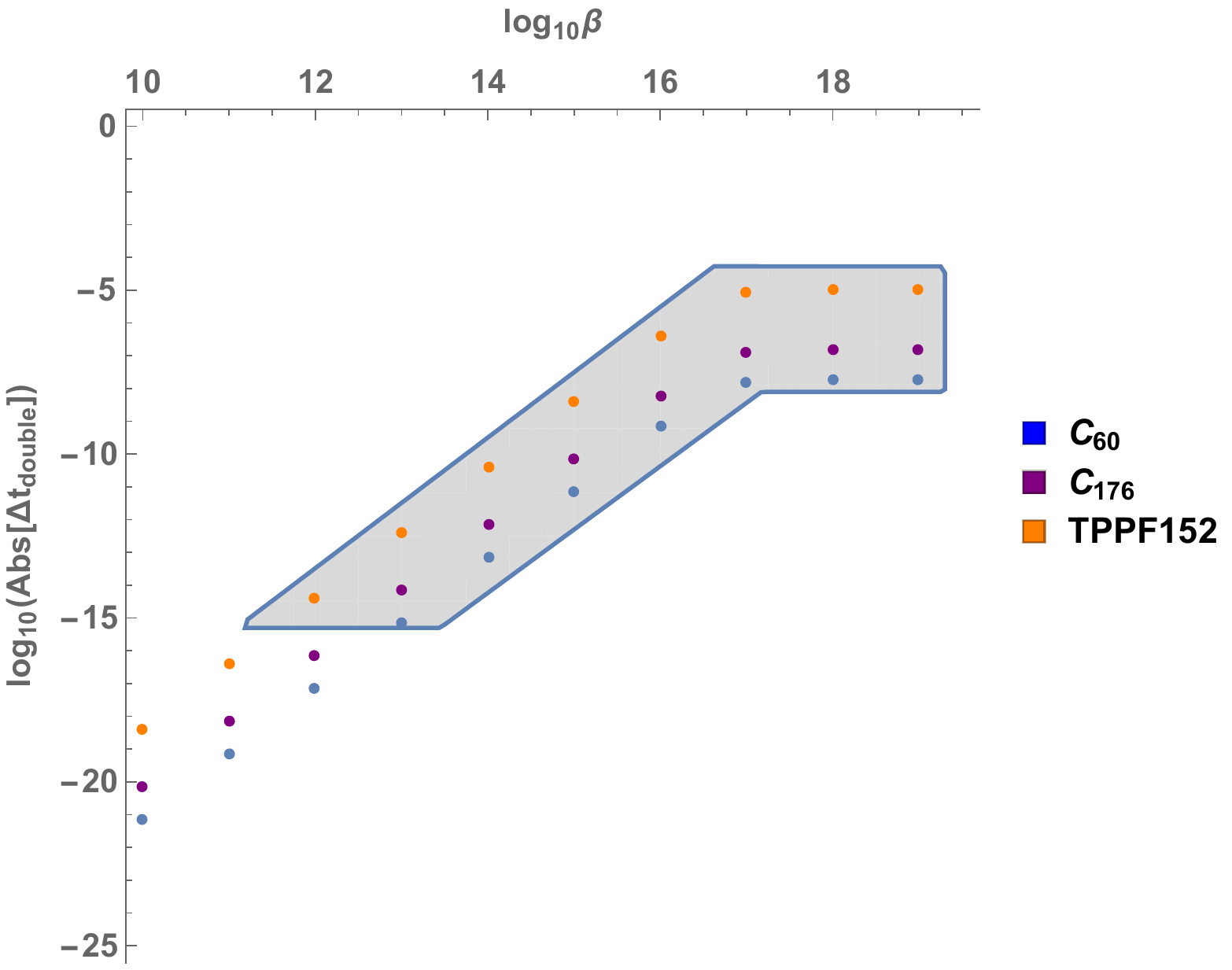}
\caption{$\log$ - $\log$ plots between the GUP parameter $\tilde{\beta}$ and the doubling time difference $\Delta t_{\text{double}}^{\text{KMM}} = t_{\text{double}}^{\text{KMM}}(\text{GUP}) - t_{\text{double}} (\text{HUP})$ between the GUP and HUP time evolution for (a) $C_{60}$ molecule (the lower plot), (b) $C_{176}$ molecule (the middle plot) and (c) the large organic molecule TPPF152 (the upper plot). The $X$ axis represents the GUP parameter and the $Y$ axis represents the difference between GUP and HUP doubling times. The shaded region indicates the region of parameter space that can be probed by the above molecular wave-packets with an atomic clock of maximum precision $10^{-15}~s$.}
\label{kmm-llplot}
\end{figure}

In Fig. \ref{kmm-llplot} we present a plot for the doubling time differences predicted by the KMM GUP, analogous to that of Fig. \ref{llplot} for the ADV form. The values of $\Delta t_{\text{double}}^{\text{KMM}}$ are shown for the parameter space $10^{10} \leq {\tilde{\beta}} \leq 10^{19}$ since these are the values for which the time differences have significant values (that may be possible to measure in a laboratory). The shaded region represents the portion of parameter-space that could be possible to probe, given the assumption that today's best atomic clocks can resolve time differences $\sim {\cal O} (10^{-15}) ~s$. Note, however, that it is possible for the currently achievable precision to be even better than this conservative estimate. Assuming this time resolution, $C_{60}$ and $C_{176}$ molecules can probe, on the lower end, down to $\tilde{\beta} \sim {\cal O} (10^{13})$, while using TPPF152 molecules would let us probe down to $\tilde{\beta} \sim {\cal O} (10^{12})$. These values of $\tilde{\beta}$ will  constrain the coupling (the quadratic term) of KMM form $\tilde{\beta}^2 \le 10^{24}$. This is again a 15 orders of magnitude better than the bound claimed $\tilde{\beta}^2 \le 10^{39}$ (coming from the study of cold atom recoil experiment as reported in \cite{pra}).

\section{Conclusions and Discussions}
\label{con}

We have introduced a novel approach and, to some extent, established the fact that studying the dispersion of free wave-packets might lead to an indirect evidence for the long anticipated minimal length scale in Nature. Our result here is based on the possibility that HUP should be replaced by a GUP in presence of the minimal length. Nonetheless, it is very important to stress that our approach is quite general and independent of the specific manner in which the commutator bracket has to be modified. This specific study, based on two popular choices given by the ADV form \eqref{gamma} and the KMM form \eqref{kmm-gamma}, has several interesting outcomes which we enlist below.

(i) For both forms of GUP the deformations have brought a rich distributional consequence on the expansion rate of free wave-packets. The rate of dispersion not only depends on the initial uncertainty and standard deviation (in position and momentum)  but also on the higher order moments in momentum space (such as skewness and kurtosis). In addition, it also depends on the initial momentum of the wave-packet. 

(ii) The minimal width of a free wave-packet is modified - for the ADV form it is normally squeezed in position space while for the KMM form it is further widened.
 
(iii) We have shown that by measuring the ``doubling time'', that is the time in which a free, minimal width wave-packet doubles its size, we may get important clues on the minimal length scale. The difference between the doubling times of HUP and GUP based predictions may well be in the detectable range if we use highly precise atomic clocks and measure the broadening rates of molecular wavepackets.

(iv) This difference in broadening time is more for massive molecular wave-packets in comparison with the wave-packets representing smallest objects like electrons. Large organic molecule (such as TPPF152), ``buckyball'' (such as $C_{60}$, $C_{176}$) wave-packets may be useful on verifying or falsifying the GUP proposals. 

(v) For the ADV form, in the absence of detecting any  difference for doubling time with an atomic clock of precision level $10^{-15}~s$,  with $C_{60}$, we can better the best existing upper bound on $\alpha$($< 10^{10}$) by one order of magnitude,  for $C_{176}$ by two orders of magnitude, and for TPPF152 by four orders of magnitudes ($\alpha<10^{6}$). This bound can be further sharpened by using atomic clocks more precise than femto-second.

(vi)  For the KMM form, use of TPPF152 and clock of femto-second precision can provide an upper limit $\tilde{\beta}^2 \le 10^{24}$.  This is an improvement by 15 orders of magnitude of the coupling $\tilde{\beta}^2$ obtained from the cold atom recoil experiments \cite{pra}.

(vii) There are two ways to improve the numbers presented here and to reach even closer to testing the GUP theory. One of them is to consider larger and heavier molecular wave-packets and the other is to come up with new atomic clocks which can measure the time difference even beyond a femto-second. 

We want to stress that coming up with an experiment to test our results might not be impossible in near future, especially because of the remarkable progress that has been achieved to test the superposition principle with increasingly massive molecular wave-packets \cite{LOM}. Perhaps, an experiment in our context will be easier to conduct since the wave-packet does not pass through the double slit, rather, it only needs to be set free until it doubles its size. 

Finally, we want to add some general remarks on the relativistic extension of the GUP setting. It should be noted that our work in this paper is based on the non-relativistic quantum mechanics as the molecular wave-packets are highly non-relativistic objects. If someone wants to study an ultra-high-energy particle with relativistic velocity this approach may not be as useful. One would need to reformulate the question in terms of fundamental quantum fields, their particle excitation etc. One has to also rethink about the physical variables that must be built from the quantum field and its derivatives which might be considered  as observables. Width of the wave-packet would not have a meaningful usage there. Therefore, one naturally asks the question to oneself that, should, if at all, GUP be used in such a context with relativistic particles? A satisfactory answer to this question can be found in an important work by Magueijo and Smolin \cite{DSR3}, where a concise picture of the GUP modification, in a relativistic setting was outlined. It was shown that the two forms of GUP (which we used here) can be derived by satisfying following five criteria - (i) validity of relativity in inertial frames, (ii) an invariant energy/length scale at Planck value, (iii) a varying speed of light at higher energies, (iv) a modified dispersion relation at higher energy (inspired by the Ultra-High-Energy-Cosmic-Ray anomaly), and (iv) that the theory should have a maximum momentum. The idea was to keep the principle of the relativity of inertial frames by modifying the laws by which energy/momenta measured by various inertial observers are related to each other. The only possibility to achieve all of these conditions was argued to be achieved by a non-linear action of the ordinary Lorentz group (in momentum space) on the states of the theory. The appearance of the Lorentz group is quite interesting and renders the theory as Lorentz invariant. In fact a naive judgement that the minimal length scale breaks the Lorentz invariance is false, and recent works on ``modular spaces'' further hints this possibility \cite{modular}. However, the isssue of fully relativistic approach is beyond the scope of this paper since our aim is to use low-energy atomic-molecular experiments in the search of fundamental length scale.

\section*{Acknowledgements}
CV thanks Fermilab for allowing him the office space and research facilities during his stay where a major part of the research work was completed. Research of SKM is supported by the {\it Consejo Nacional de Ciencia y Tecnolog\'ia} (CONACyT) Project No. CB17-18/A1-S-33440 (Mexico).


\end{document}